# Lord of the Ring Gratings:
# Their use as image plane filters for coronagraphy

François Hénault
Institut de Planétologie et d'Astrophysique de Grenoble
Université Grenoble-Alpes, Centre National de la Recherche Scientifique
B.P. 53, 38041 Grenoble – France

## ABSTRACT

Coronagraphy is an efficient technique for identifying and characterizing extra-solar planets orbiting in the habitable zone of their parent star. An important family of coronagraphs is based on amplitude or phase filters placed at an intermediate image plane of the optical system, spreading starlight outside of the so-called "Lyot stop" located at the exit pupil plane of the instrument. This article explores the potential of circular amplitude and phase gratings employed as image plane coronagraph filters. It presents a theoretical analysis of the simplest case of an amplitude circular grating and introduces an inversion paradigm with respect to classical Lyot coronagraph, by exchanging its image and pupil masks. Various types of circular gratings are considered and their performance is evaluated with the help of numerical simulations. The most promising solutions are presented. We conclude that high attenuation ratios of the starlight are feasible, provided that the system has been carefully optimized.

## 1 INTRODUCTION

In the forthcoming years, coronagraphy will probably be the most efficient technique for identifying and characterizing extra-solar planets orbiting in the habitable zone of their parent star, as already demonstrated by the ground instruments SPHERE [1] and GPI [2] in operation on the Very Large Telescope (VLT) and Gemini South telescope respectively. Future space observatories like WFIRST-AFTA [3] are also in construction. The present paper is especially focused on Lyot type coronagraphs (LTC), where an amplitude or phase filter is located at an intermediate image plane of the optical system, and diffracts starlight outside of the diameter of a "Lyot stop" placed at the exit pupil of the system, before finally focusing the beam at the image plane of the instrument. Among this family, one will find the historical Lyot coronagraph where the image filtering is achieved with a small-size central blocking mask [4]. Later came other designs named Phase mask coronagraph (PMC) that make use of nulling phase jumps at the focal plane of the telescope. But so far, it seems that little attention was paid to circular amplitude or phase gratings applied to coronagraphy [5]. The scope of this paper is then to explore this vast and little-known realm, including description of the design, numerical simulations and discussion about the achievable performance.

## 2 DESIGN DESCRIPTION

### 2.1 General principle

Let us consider the case of a coronagraphic telescope of focal length $F$ and diameter $D = 2R$, equipped with a circular grating located in its image plane. The employed coordinate systems are sketched in Figure 1. A brief summary of the scientific notations and of the mathematical development of the intensity distributions formed in the Lyot stop plane is given into the Appendix.

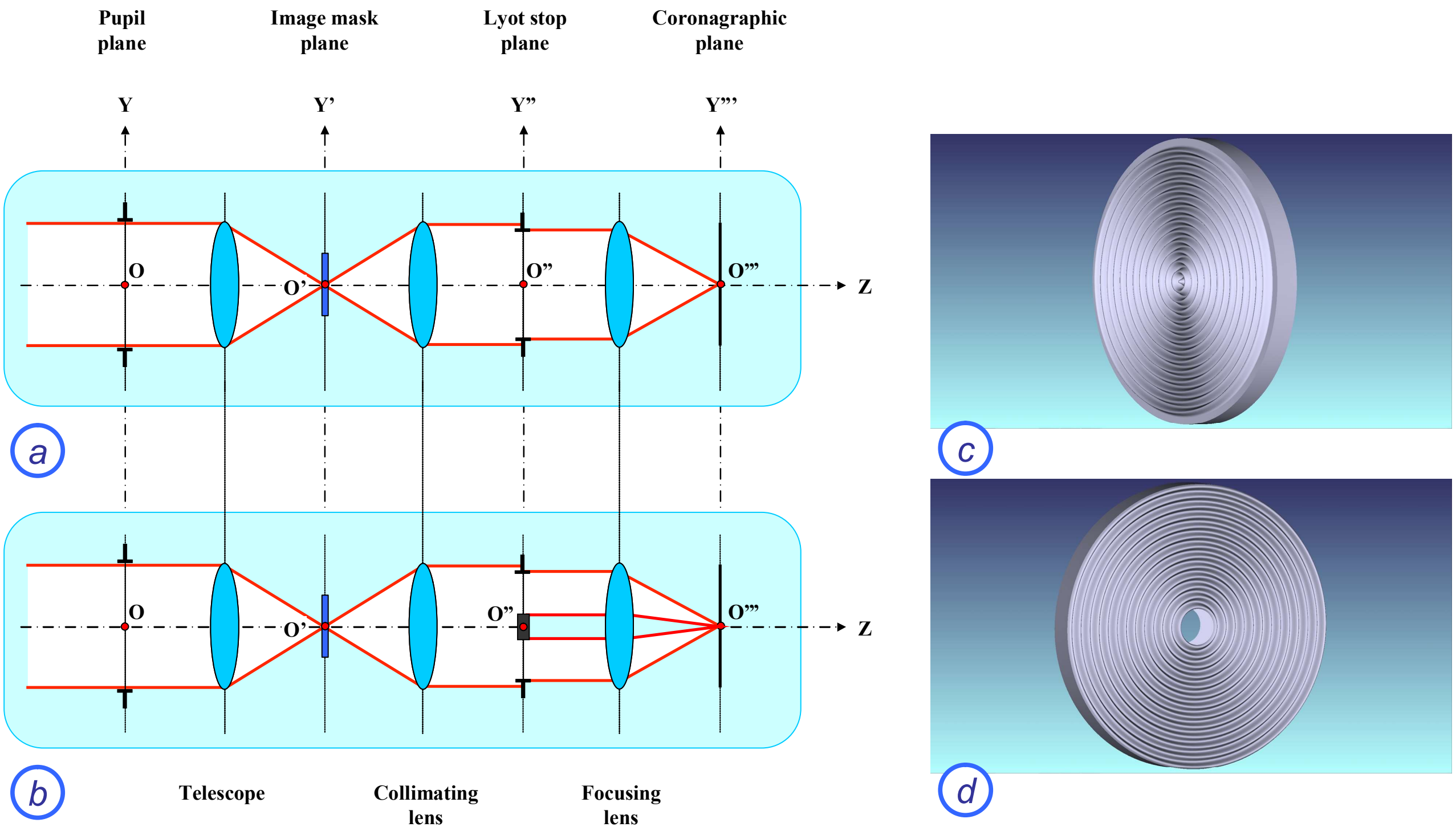

Figure 1: General sketch of a LTC (*a*) and of the modified layout for circular grating filters (*b*). (*c*–*d*): Schematic views of a transmitting circular phase grating.

This analytical development demonstrates that the intensity distribution in the Lyot stop plane reduces to a bright central spot, if and only if one unique condition is fulfilled [6]:

$$f' = R/\lambda F = 1/2\lambda N \tag{1}$$

where $f'$ denotes the spatial frequency of the grating, $\lambda$ is the wavelength and $N$ the telescope aperture number. It follows that the spatial period of the grating must be equal to $p' = 2\lambda F/D = 2\lambda N$, thus being matched to the cut-off spatial frequency of the telescope.

Eq. 1 is at the origin of an "inversion paradigm" that results in defining unconventional Lyot stops with central obscuration. Practically, it consists in blocking the central area of the Lyot stop (see Figure 1-*b*) instead of the image plane as in the classical Lyot coronagraph (Figure 1-*a*).

## 2.2 Panorama of amplitude and phase gratings

There actually exists a wide variety of circular gratings: firstly, they can either be Circular amplitude gratings (CAG) whose diffracting pattern is coded into a gray scale transmission map varying between zero and unity, or transparent Circular phase gratings (CPG) introducing periodic phase distributions overall their surface. Ruled gratings belong to the first category and Volume phase holographic gratings (VPHG) to the second one. The latter looks more interesting because of a higher potential transmission, on the one hand, and the amplitude $\phi'$ of the phase variations represents an additional design parameter, on the other hand. Next, the diffractive pattern can either be binary or continuous. In the latter case, various grating profiles may be designed, for example cosine or sine profiles, saw tooth profile and triangle profile. All of them can be "phase-shifted" by a quantity $\varphi$ along the radial axis, whether they are binary or not. Binary CAG can be made of either equally-spaced black and white alternating strips, or of unequal slit widths. Continuous CAG functions can also be raised to the power of an integer or real number $\xi$ with resulting grating transmission gains or losses. Finally, the axis symmetric geometry of CAG and CPG can be replaced with spiral geometry [6]. It follows that there exists a quasi-infinite number of solutions, especially when considering that the previous options can be combined with each other. Such a wide panorama is illustrated in a humoristic way in Figure 2. For the sake of illustration, Figure 3 displays nine pictures of intensity distributions produced in the Lyot stop plane by a few typical ring gratings. It is quite remarkable that in all cases the intensity distribution exhibits a bright central spot whatever is the nature of the circular grating (amplitude or phase, grooves shape, binary or not, etc.), as long as the condition defined by Eq. 1 is respected. One may also note the presence of a faint diffraction halo surrounding the central peak, which will stand for the main limitation of the achievable attenuation factors. The numerical simulations presented in the next section will be helpful to better characterizing this effect.

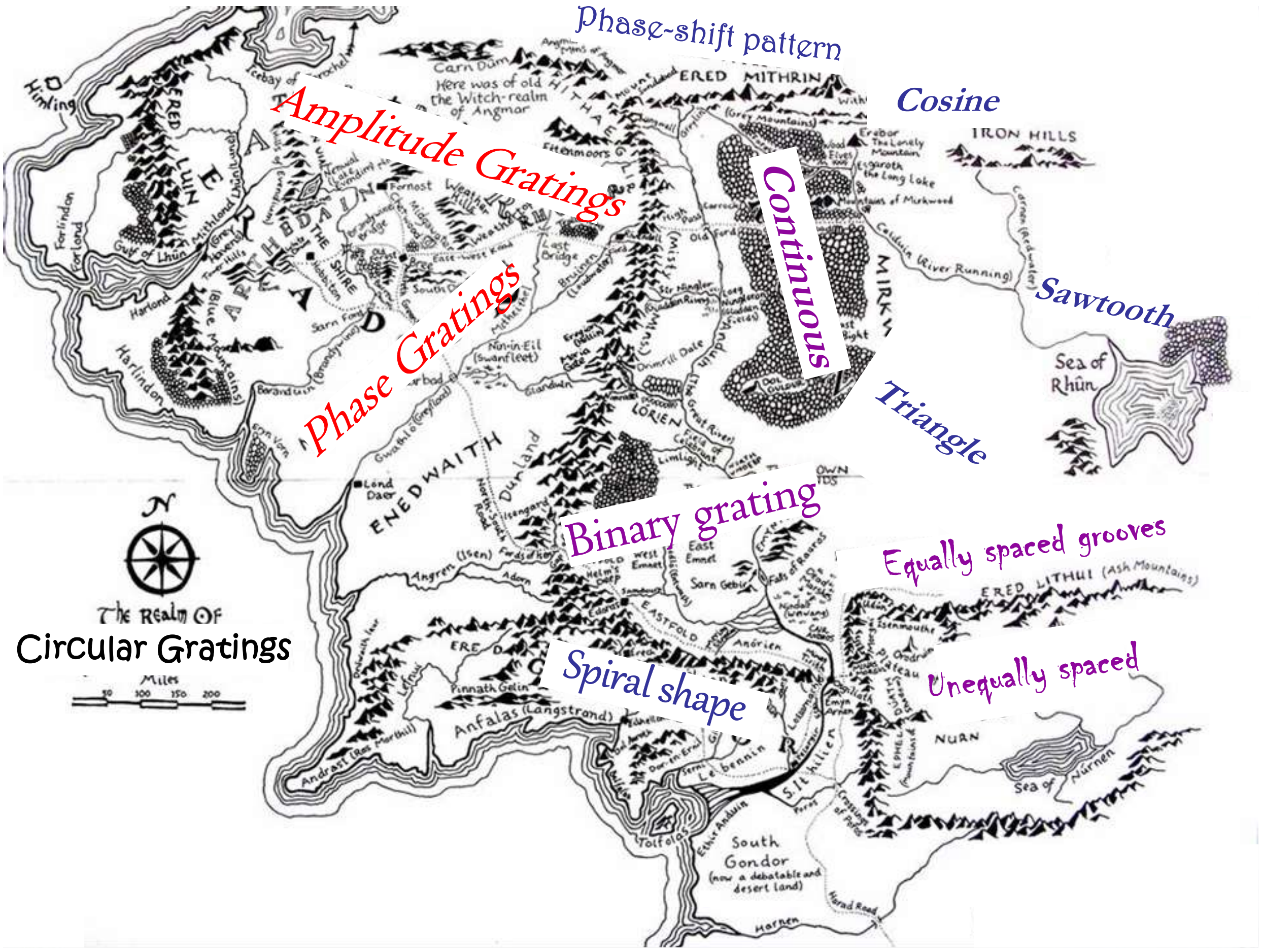

Figure 2: Tentative map of the realm of circular gratings.

## 3 NUMERICAL SIMULATIONS

### 3.1 Numerical model

Herein are presented numerical simulations of the intensity distributions produced into the Lyot stop plane, and of the resulting Nulled Spread Function (NSF) into the final image plane of the coronagraph. They are achieved with a cosine and spiral-shaped GPG. It is assumed that the wavelength $\lambda$, telescope diameter $D$, and telescope focal length $F$ are equal to 0.5 µm, 10 m and 200 m respectively. The employed numerical model has been described in Ref. [6]. The numerical results are summarized in Table 1 in terms of maximal intensity peak at the Lyot stop plane and extinction ratio at the final image plane of the coronagraph. They are illustrated in Figure 4.

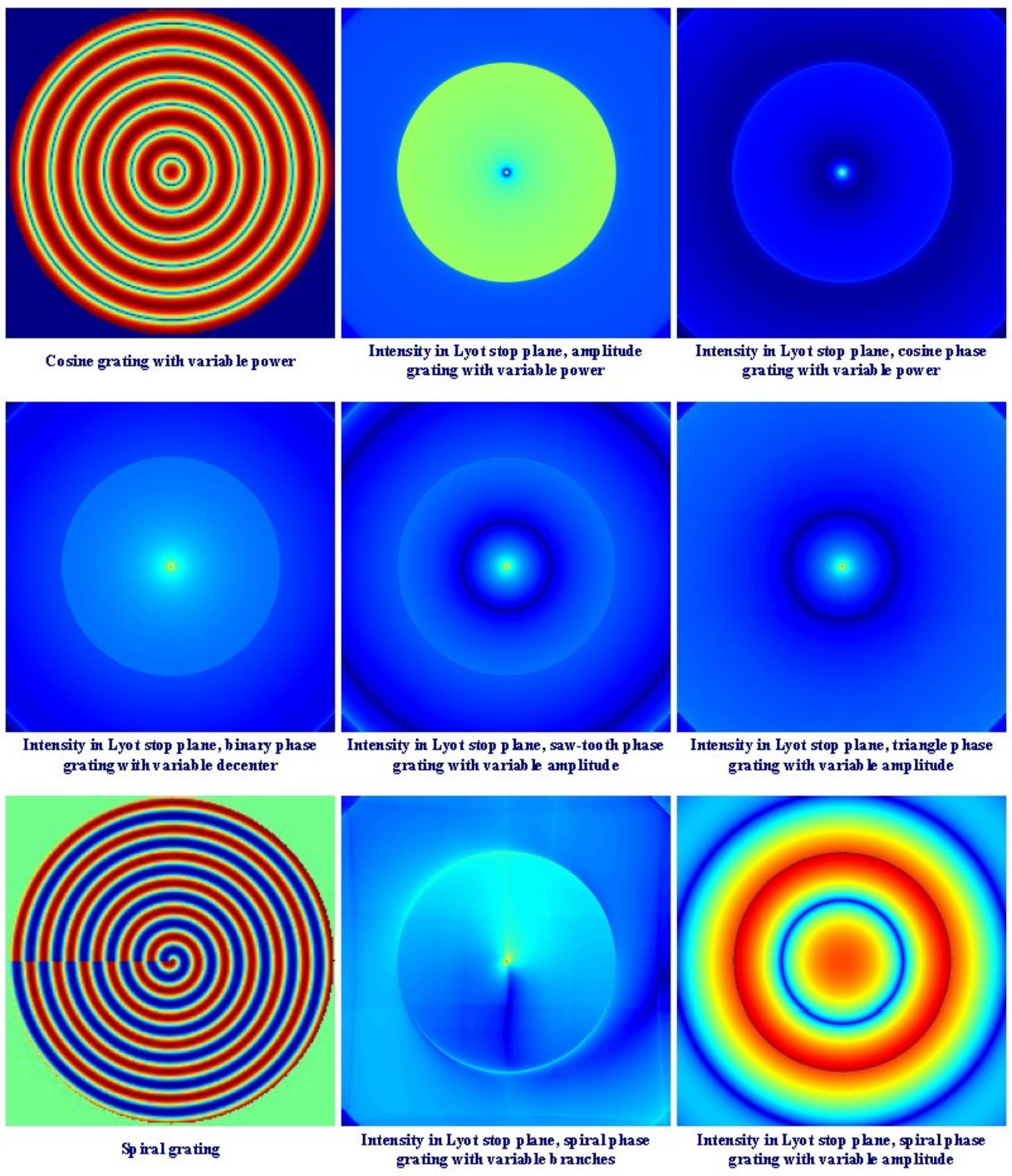

Figure 3: Illustrating the intensity distributions produced in the Lyot stop plane by nine "Nazgul" ring gratings in logarithmic scales. Additional animated illustrations can be seen at [7].

Firstly, the Figure shows two customized Lyot stop masks (black and white maps in Figure 4-*a* and *b*) that were optimized for the grating shapes and parameters in order to minimizing the star nulling ratio, on the one hand, and to maximizing the off-axis planet transmission efficiency, on the other hand. It can be seen that the light emitted by the extra-solar planet is diffracted by the grating over a much larger area than the classical circular Lyot stop (indicated by blue circles in Figure 4-*c* and *d*, and red circles in Figure 4-*e* and *f*). Hence the Lyot stops were enlarged along the horizontal axis accordingly, and their parameters are indicated in Table 1. Finally, the Figure 4-*g* and *h* show false-color views of the NSF formed at the final image plane in logarithmic scale.

Table 1: Numerical results obtained for two enhanced CPG designs.

| Grating type | Grating profile | Phase amplitude | Filter obscuration ratio | Maximal peak at Lyot stop | Lyot stop threshold | Lyot stop dimensions y" / x" | Central star null |
|---|---|---|---|---|---|---|---|
| CPG | Cosine | $\pm 3.89\,\pi$ | 0.035 | 184.5 | 6.0E-05 | 5 / 1.2 | 1.2E-04 |
| CPG | Spiral, charge $s = 3$ | $\pm 0.27\,\pi$ | 0.113 | 14.5 | 1.8E-05 | 5 / 1.2 | 1.9E-06 |

### 3.2 Wavelength dependence

The variations of the nulling ratio at wavelengths $\lambda$' differing from the nominal one is of major importance. Then the numerical model was used to compute the nulling rates over an extended spectral bandwidth of 20%, i.e. $|(\lambda'-\lambda)/\lambda| < 0.1$. Plots of the results in logarithmic scale are shown on the left side of Figure 5 for the cosine (blue curve) and spiral cases (red curve). Additionally, the grey curve shows the wavelength sensitivity of a vortex coronagraph of topological charge 2, one of the most popular. It is found that the null ratio variations of the cosine and spiral-shaped CPG are less than one order of magnitude. In opposition, the vortex coronagraph could only attain such performance over a spectral bandwidth about 1%. These results give a clear advantage to the CPG with respect to other types of PMCs. It also has to be noted that such low wavelength sensitivity only involves the intrinsic properties of the grating.

### 3.3 Dependence to planet field angle

A satisfactory coronagraph design should not only provide high attenuation factors and low wavelength sensitivity, but also good transmission to the off-axis extra-solar planet orbiting around its star. The transmission estimates are plotted in logarithmic scale as function of a field angle ranging from 0 to 100 $\lambda/D$ on the right side of Figure 5. They reveal moderate transmission performance, varying from 20% to 40% at most. These curves are also useful to defining the Inner working angle (IWA) of the coronagraph, i.e. the angular separation for which the planet transmission is equal to half of its maximal value. It can be seen that the IWAs are significantly larger than those achieved by other types of PMC, typically about 5 $\lambda$/D and 25 $\lambda/D$ for the cosine and spiral-shaped CPGs respectively. Such modest performance numbers in terms of transmission and IWA actually are the main disadvantages of the CPG coronagraph, to be counterbalanced with its much larger spectral bandwidth.

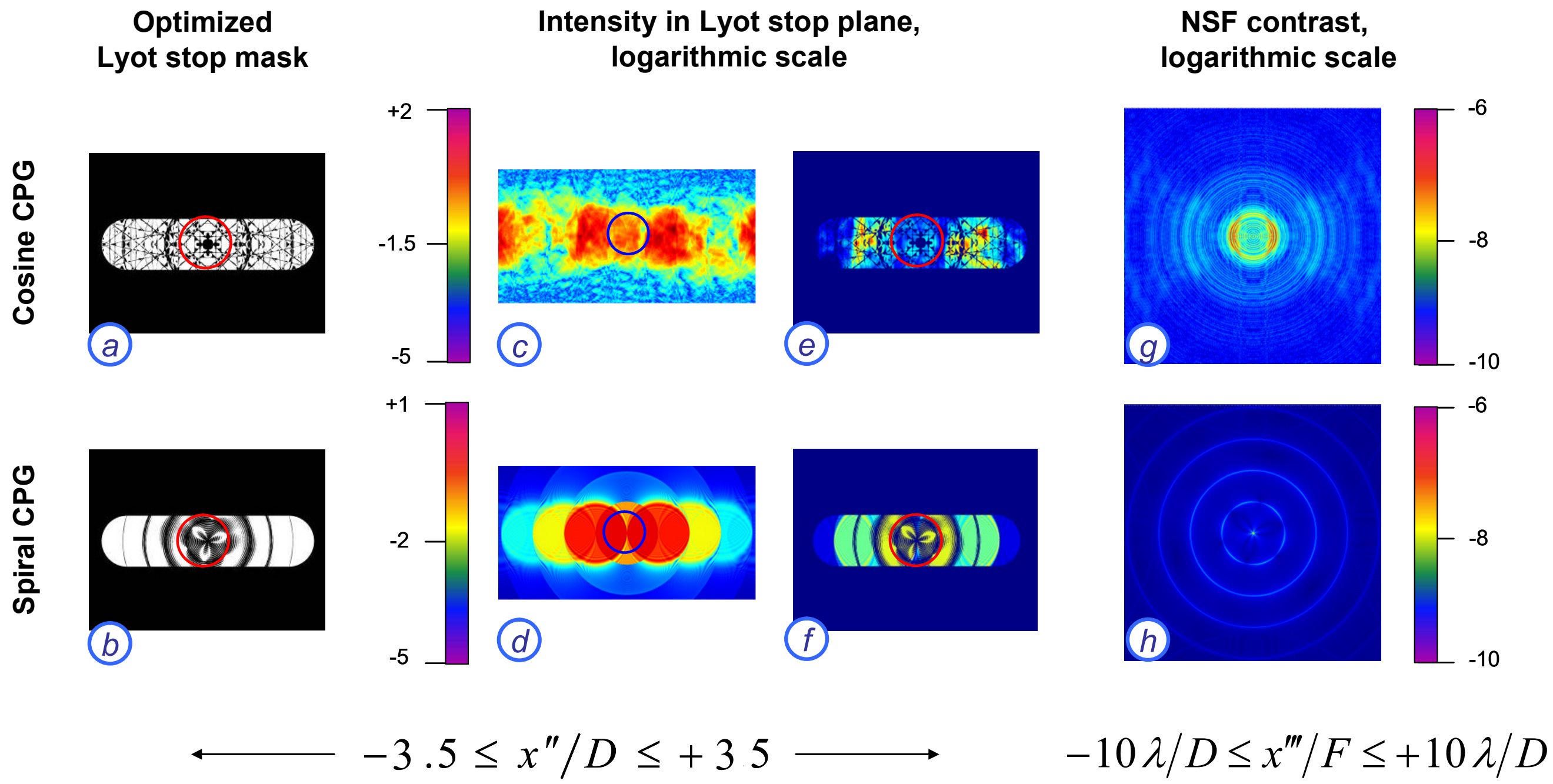

Figure 4: Optimized CPG solutions. (*a–b*): B black and white maps of the optimized Lyot stop masks for the cosine and spiral-shaped CPG cases. (*c–d*): False-color views of the intensities formed into the Lyot stop plane, unmasked (logarithmic scale). (*e–f*): Same illustrations of the intensities after being blocked by the Lyot stop. (*g–h*): False-color views of the resulting NSF into the final image plane. The diameter of the classical circular Lyot stop is indicated by red or blue circles.

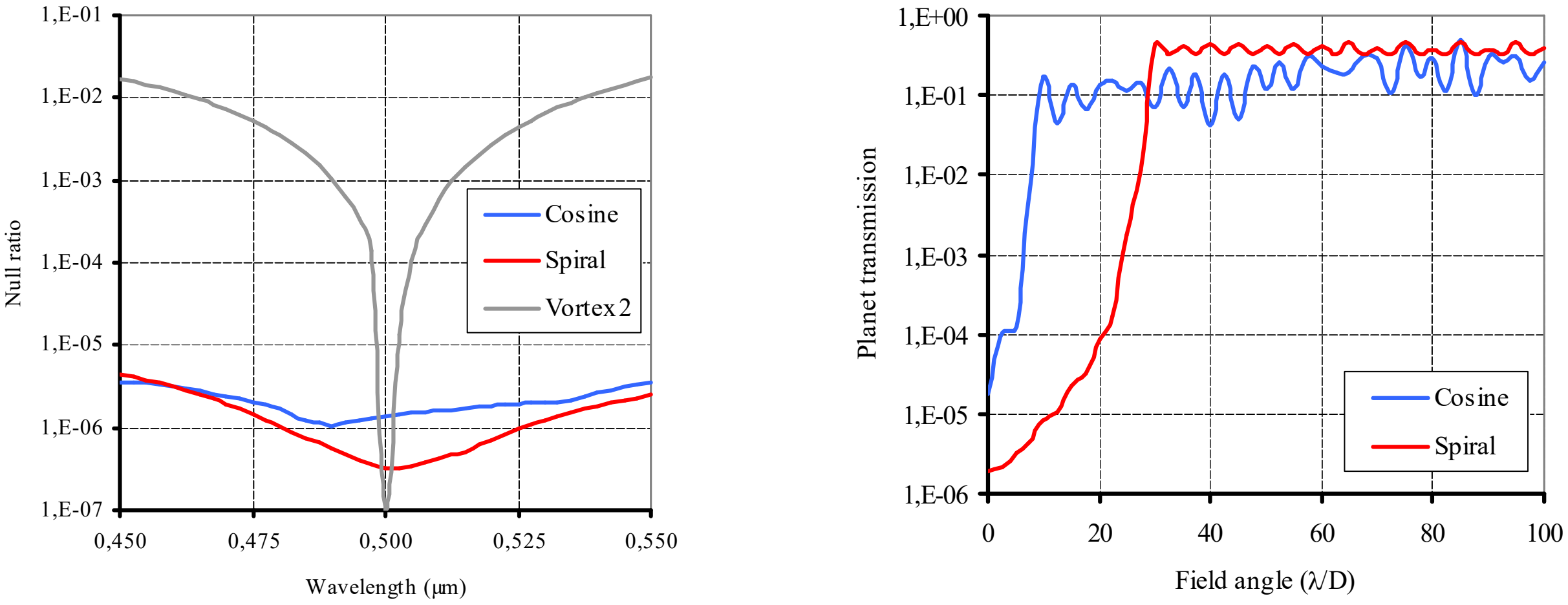

Figure 5: Plots of nulling rates in an extended spectral bandwidth of 20% (left) and transmission curves achieved by the cosine and spiral gratings (right).

# 4 CONCLUSION

Arriving at the end of our first journey through the realm of Ring Gratings applied to coronagraphy, it is already possible to draw some conclusions. Firstly, when the spatial frequency of the grating is properly matched to the exit pupil of the optical system, such amplitude or phase gratings generate a bright central spot at the Lyot stop plane. This is at the origin of an inversion paradigm that consists in blocking the central area of the Lyot stop instead of the image plane as made for classical Lyot coronagraphs. There exists a wide variety of grating types and shapes, among which sawtooth, triangular, binary, spiral, or multiple combinations of them. Such a large number of cases can only be handled with numerical simulations.

Secondly, the performance of a CPG coronagraph can be improved by optimizing both the grating parameters and Lyot stop transmission mask. It leads to the conclusion that a spiral phase grating probably stands for the best compromise between the star nulling ratio and planet transmission.

Thirdly; the GPG coronagraph has been compared with the vortex coronagraph that is considered as one of the most efficient nowadays. It turns out that the CPG exhibits worse performance in terms of planet transmission and inner working angle, on the one hand, but shows the unique ability to operating over a much larger spectral bandwidth, on the other hand. It can be concluded that both designs are very complementary.

## APPENDIX. THEORETICAL ANALYSIS

This Appendix gives a brief summary of the main employed scientific notations and of the analytical development of the intensity distribution $I''(\rho'',\theta'')$ formed in the Lyot stop plane, for the simplest case of a cosine-shaped amplitude transmission grating (CAG). From Ref. [8] the complex amplitude $A'(\rho',\theta')$ diffracted by a non aberrant telescope pupil into the image plane writes in polar coordinates:

$$A'(\rho',\theta') = K \int_0^{2\pi} \int_0^{+\infty} B_D(\rho) \exp[-2i\pi\rho\rho' \cos(\theta'-\theta)/\lambda F]\, \rho d\rho d\theta \,. \tag{A1}$$

where K is a normalization factor. Multiplying this complex amplitude with the CAG transmission function equal $0.5 + 0.5\cos(2\pi f'\rho')$ and taking the inverse Fourier transform of the result leads to the analytical expression of the complex amplitude distribution $A''(\rho'',\theta'')$ at the Lyot stop plane:

$$A''(\rho'',\theta'') = \pi^2 \frac{R^2}{\lambda F} \int_0^{+\infty} J_0(2\pi\rho''\rho'/\lambda F) J_1(2\pi R\rho'/\lambda F) [1+\cos(2\pi f'\rho')]\, d\rho' \,, \tag{A2}$$

Omitting the multiplying factor $\pi^2 R^2/\lambda F$ , $A''(\rho'',\theta'')$ can be expressed as the sum of two terms:

$$A''(\rho'',\theta'') = A''_0(\rho'',\theta'') + A''_C(\rho'',\theta''),$$

where: $$A''_0(\rho'',\theta'') = \int_0^{+\infty} J_0(2\pi\rho''\rho'/\lambda F) J_1(2\pi R\rho'/\lambda F)\, d\rho' \tag{A3}$$

and $$A''_C(\rho'',\theta'') = \int_0^{+\infty} {}_0(2\pi\rho''\rho'/\lambda F) J_1(2\pi R\rho'/\lambda F) \exp(2i\pi f'\rho')\, d\rho' \,.$$

Referring to mathematical textbooks [9-10], the analytical expression of $A''_0(\rho'',\theta'')$ reduces to bias terms. It is also found that:

$$A''_C(r'',\theta'') = \frac{i}{2\pi}\int_0^{2\pi} \sin(t)\, G(r'',t)\, dt$$

with: $$G(r'',t) = R\int_0^{+\infty} J_0(2\pi r'' r')\exp[2i\pi(b''+\cos t)r']dr' ,\qquad \text{(A4)}$$

and using non dimensional, reduced notations $r' = \rho'/R$, $r'' = \rho''/R$ and $b'' = 2\lambda N f'$

The analytic expression of $G(r'',t)$ derived from Refs. [9-10] is then:

$$\begin{cases} G(r'',t) = \dfrac{iR}{\sqrt{(b''+\cos t)^2 - r''^2}} & \text{if} \quad r'' < b''+\cos t \\ G(r'',t) = 0 \ \text{ or } \infty & \text{if} \quad r'' = b''+\cos t \\ G(r'',t) = \dfrac{R}{\sqrt{r''^2 - (b''+\cos t)^2}} & \text{if} \quad r'' > b''+\cos t\,. \end{cases} \qquad \text{(A5)}$$

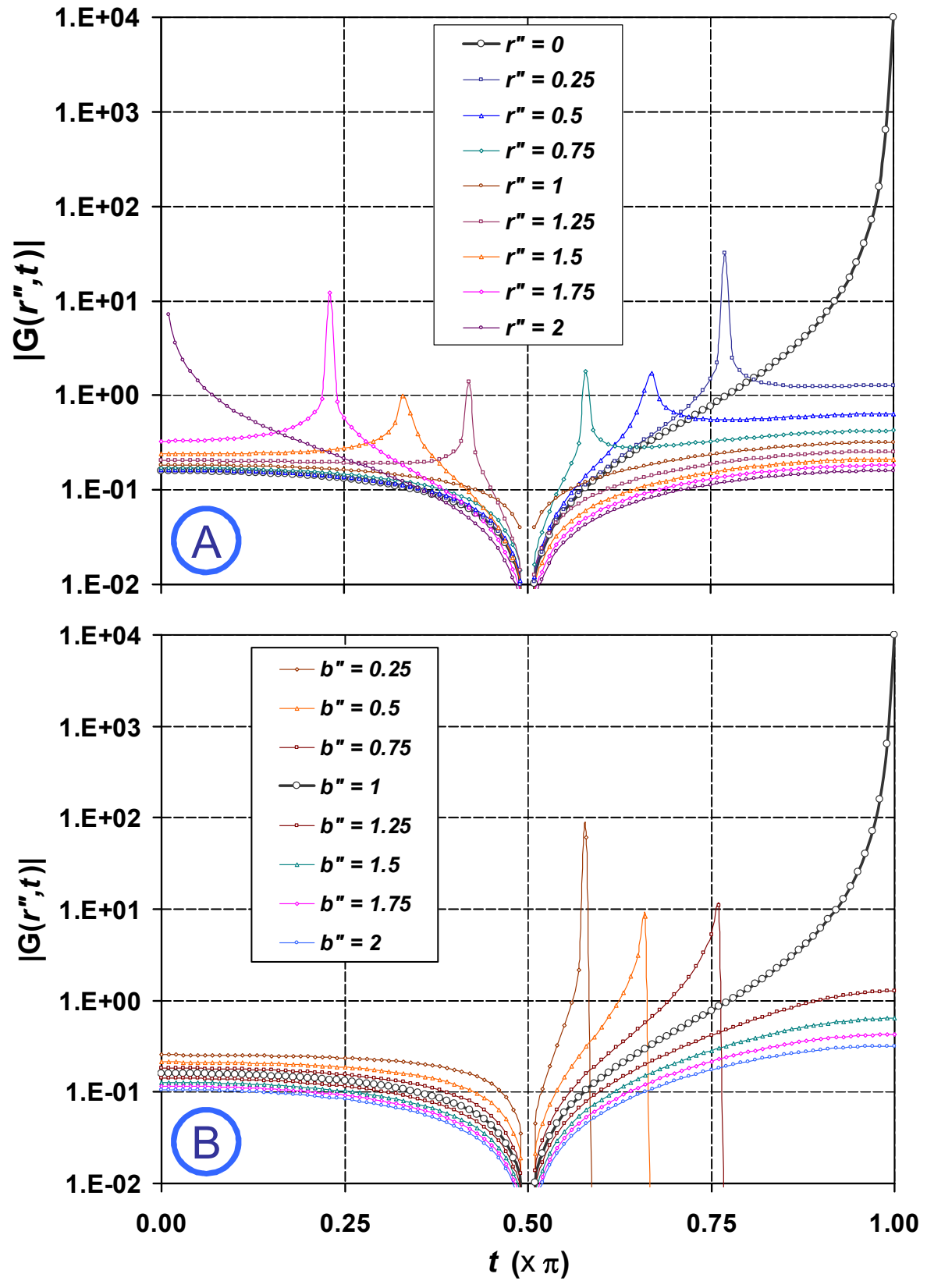

Figure A1: (A): Plots of function $G(r'',t)$ in logarithmic scale as function of *t* in the interval [0–π] at different polar coordinates $r''$ when $b'' = 1$. (B) Plots of $G(r'',t)$ at different grating spatial frequencies $b''$ when $r'' = 0$. Curves in the interval [π –2π] can be obtained *via* vertical symmetry.

Plots of $G(r'',t)$ as function of the reduced parameters $r''$ and $b''$ are shown in Figure A1. Its integration bounds are determined by the conditions $r'' < b'' + \cos t$ and $r'' > b'' + \cos t$, leading to an expression of $A''_C(r'',\theta'')$ as:

$$A''_C(r'',\theta'') = \frac{1}{2\pi} \int_{+\arccos(r''-b'')}^{2\pi-\arccos(r''-b'')} \frac{\cos t \, dt}{\sqrt{r''^2 - (b'' + \cos t)^2}} \; . \tag{A6}$$

Finally adding the bias terms $A''_0(r'',\theta'')$ to Eqs. A6, the analytic expression of the amplitude distribution in the Lyot stop plane writes as a function of the reduced polar coordinate $r''$:

$$\begin{cases} A''(r'') = \dfrac{1}{2\pi}\left\{1 + \displaystyle\int_{+\arccos(r''-b'')}^{2\pi-\arccos(r''-b'')} \frac{\cos t \, dt}{\sqrt{r''^2 - (b''+\cos t)^2}}\right\} & \text{if} \quad r'' < 1, \\ A''(r'') = \dfrac{1}{2\pi} \displaystyle\int_{+\arccos(r''-b'')}^{2\pi-\arccos(r''-b'')} \frac{\cos t \, dt}{\sqrt{r''^2 - (b''+\cos t)^2}} & \text{if} \quad r'' > 1. \end{cases} \tag{A7}$$

Plots of the complex amplitude $A''(r'')$ and of the resulting intensity $I''(r'') = |A''(r'')|^2$ in the Lyot stop plane are reproduced in Figure A2 for the most interesting case when the reduced parameters $r''$ and $b''$ are equal to 0 and 1 respectively. In that case the complex amplitude $A''(0)$ is indeterminate mathematically, but deemed to be equal to complex infinity $-i\infty$. This is the heart of the necessary condition to forming to a bright spot at the centre of the Lyot stop, as defined in the main text by Eq. 1. The Figure also evidences the presence of the faint diffraction halo surrounding the central peak.

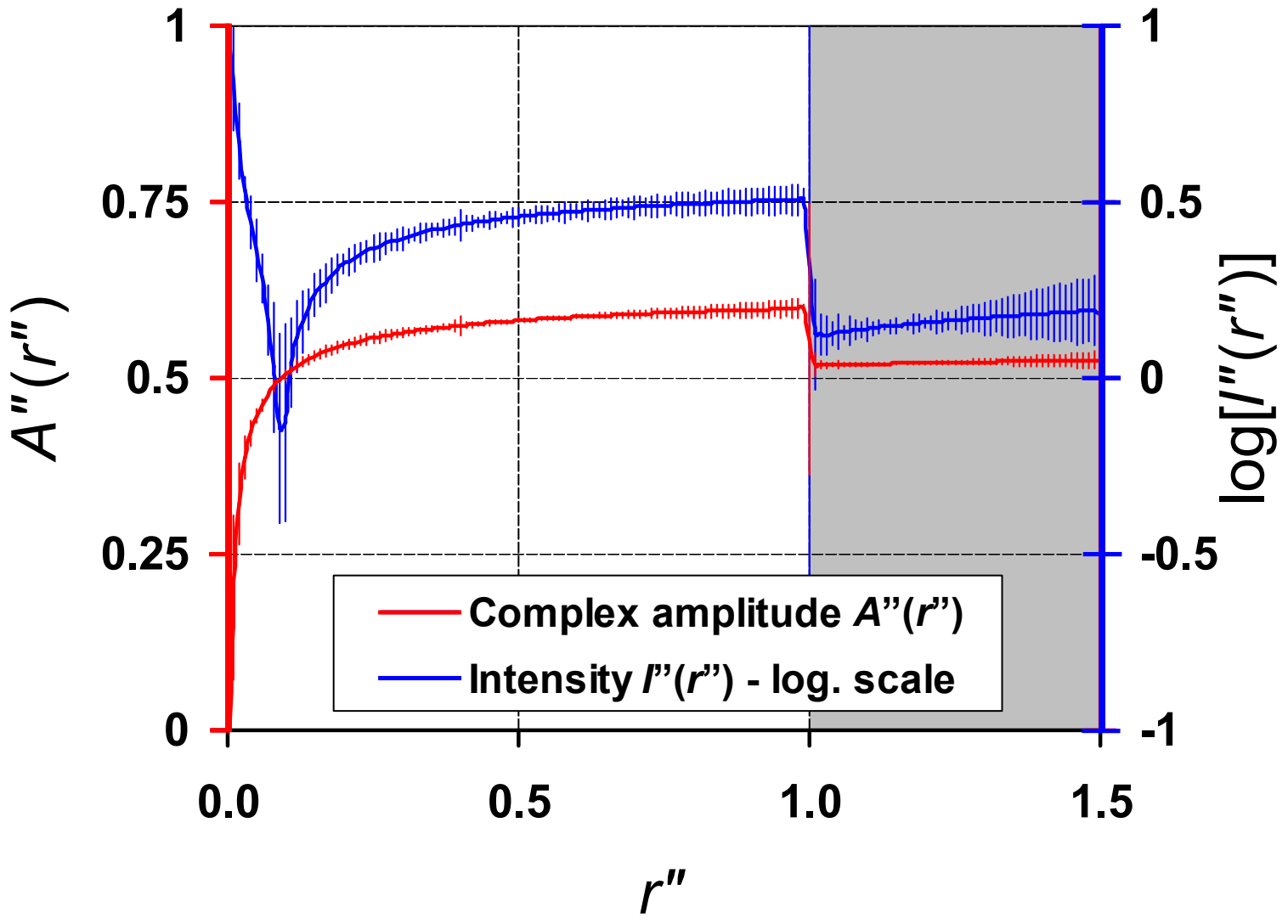

Figure A2: Plots of the complex amplitude $A''(r'')$ as function of $r''$ (red curve) and of the intensity $I''(r'')$ formed in the Lyot stop plane (blue curve in logarithmic scale).